\documentclass[aps,pra,twocolumn]{revtex4}

\usepackage{graphicx}
\usepackage{subfigure}
\usepackage{amsmath}
\usepackage{amssymb}
\usepackage{mathrsfs}
\usepackage{hyperref}
\hypersetup{colorlinks=true, linkcolor=black, citecolor=black, urlcolor=black}

\begin{document}
\title{Asymptotically perfect efficient quantum state transfer across uniform 
chains with two impurities}
\author{Xining Chen}
\author{Robert Mereau}
\author{David L.~Feder}
\email[Corresponding author: ]{dfeder@ucalgary.ca}
\affiliation{Institute for Quantum Science and Technology, 
Department of Physics and Astronomy,
University of Calgary, Calgary, Alberta, Canada T2N 1N4}
\date{\today}
\begin{abstract}

The ability to transfer quantum information from one location to another with
high fidelity is of central importance to quantum information science. 
Unfortunately for the simplest system of a uniform chain (a spin chain or a 
particle in a one-dimensional lattice), the state transfer time grows 
exponentially in the chain length $N$ at fixed fidelity. In this work we show 
that the addition of an impurity near each endpoint, coupled to the uniform 
chain with strength $w$, is sufficient to ensure efficient and high-fidelity
state transfer. An eigenstate localized in the vicinity of the impurity can be 
tuned into resonance with chain extended states by tuning
$w(N)\propto N^{1/2}$; the resulting avoided crossing yields resonant 
eigenstates with large amplitudes on the chain endpoints and approximately 
equidistant eigenvalues. The state transfer time scales as $t\propto N^{3/2}$ 
and its fidelity $F$ approaches unity in the thermodynamic limit $N\to\infty$;
the error scales as $1-F\propto N^{-1}$. Thus, with the addition of two 
impurities, asymptotically perfect state transfer with a uniform chain is 
possible even in the absence of external control.


\end{abstract}
\maketitle

\section{Introduction}
\label{sec:introduction}

The ability to transfer information is crucial for digital communications.
Likewise in quantum computation and communication, the ability to efficiently 
and reliably transfer quantum information is central to both current and 
future quantum technologies~\cite{Gisin2007}. In the standard circuit model of 
quantum computation, the quantum information is encoded on localized spins with 
two or more accessible distinguishable quantum states (qubits or qudits), and 
an algorithm is effected by manipulating individual spins, performing two-spin 
operations, and making measurements~\cite{Nielsen2011}. The physical objects 
encoding the spins, for example spins in semiconductors~\cite{Awschalom2013} or 
nitrogen-vacancy centers in diamond~\cite{Doherty2013}, may be widely 
separated, requiring the development of robust quantum state transfer protocols
for spin networks~\cite{Kay2010}.

Schemes for perfect quantum state transfer (PST) in spin networks were 
developed over a decade ago~\cite{Christandl2004,Christandl2005}. Quantum 
information encoded on a spin at one end of a linear chain of nearest-neighbor 
coupled spins was found to propagate perfectly to the opposite end, as long as 
the non-uniform spin coupling coefficients could be carefully chosen. 
Subsequent work showed how to find all possible coupling constants consistent
with PST for a spin chain of arbitrary length~\cite{Wang2011,Vinet2012a}. The 
ability to adjust a large number of coupling constants is expected to be 
experimentally challenging, and unfortunately PST is not possible in uniform 
spin chains (i.e.\ where all coupling constants are identical) longer than 
three sites. This has prompted the investigation of different network 
topologies with uniform (but possibly signed) couplings that can support 
PST~\cite{Saxena2007,Facer2008,Bernasconi2008,Jafarizadeh2008,Angeles-Canul2009,Basic2009a,Basic2009b,PST2,Ge2011,PST3,Brown2013}.

An alternative strategy is to relax the assumption of perfect quantum state 
transfer, replacing it with `pretty good state transfer' 
(PGST)~\cite{Godsil-main,Godsil2012} or equivalently `almost perfect state 
transfer' (APST)~\cite{Vinet2012b}. In these cases, one sets the desired 
fidelity $F$ of the output state, and determines the time required (if any) to 
achieve it. While these equivalent concepts in the literature are frequently 
referred to as APST, we prefer to employ the term `imperfect quantum state 
transfer' (IST) to clearly distinguish the behavior from PST. In IST there 
exists a time at which the initial state at some site transfers to a different 
site with probability approaching unity to within some error. 

For uniform spin chains, IST is only possible in principle for particular 
values of the number of spins $N$~\cite{Godsil2012,Sousa2014}. For fixed 
minimum fidelity, however, the transfer time $t$ increases exponentially with 
$N$~\cite{Sousa2014}. A linear scaling of $t$ with $N$ can be achieved by 
coupling the initial and final spins only weakly to the uniform 
chains~\cite{Wojcik2005,Venuti2007,Feldman2010,Banchi2011a,Yao2011,Apollaro2012,Zwick2012,Bruderer2012,Paganelli2013,Yao2013}. Because the quantum information 
is strongly
localized in the vicinity of the chain ends at all times, this model is also 
more robust against noise than the bare uniform chain. Unfortunately, in this
model after optimizing the strength of the weak coupling parameter (which is
found to decrease like $N^{-1/6}$) and initializing the channel, the maximum 
output amplitude decreases with $N$, attaining an asymptotic value of $0.8469$ 
for $N\to\infty$~\cite{Banchi2011a}. Adding additional weak links improves the
results, but the asymptotic error in fidelity remains 
finite~\cite{Apollaro2012}. Instead applying a magnetic field in the vicinity
of the chain ends yields fidelities that approach unity in the large-$N$ limit,
but at the cost of $t$ growing exponentially with 
$N$~\cite{Casaccino2009,Linneweber2012,Paganelli2013,Lorenzo2013,Lorenzo2015}.

In this work we consider a uniform spin chain of length $N$, with sites labeled
from 1 through $N$, with the minor modification of an additional impurity spin
coupling to the spin at site $3$ and another to site $N-2$, both with coupling 
constant $w$. The value of $w$ is left as a variable to be optimized. It is 
found that there exists a value of $w\sim\sqrt{N}$ that yields a resonance 
between the quantum state localized near the impurity and extended states of 
the chain. The resulting avoided crossing yields strongly mixed eigenvectors 
with equally spaced eigenvalues and large overlaps with the chain endpoints.
Under these conditions, the fidelity for quantum state transfer from site 1 to 
site $N$ approaches unity in the thermodynamic (large-$N$) limit, with error 
$1-F\propto N^{-1}$. The system therefore exhibits `asymptotically perfect 
quantum state transfer,' a behavior previously unobserved for a spin network. 
The time is also found to scale efficiently with the chain length, 
$t_{\rm IST}\propto N^{3/2}$. In concrete terms, the output fidelity surpasses 
$F=0.9$ when $N>99$ and exceeds $F=0.99$ for $N\geq 900$.

Section~\ref{sec:IST} briefly reviews the essential characteristics of IST, 
with an emphasis on mirror-symmetric networks such as spin chains. The model of
interest in the present work is introduced in Sec.~\ref{sec:model}. The minimum
IST time found by optimizing the impurity coupling strength $w$ is found to be 
a strongly non-monotonic function of $N$. To find a functional form for $w(N)$,
the model is investigated analytically in Sec.~\ref{sec:analytics}; 
high-fidelity IST in time $t_{\rm IST}\propto N^{3/2}$ is found to occur if 
$w(N)\propto\sqrt{N}$. Section~\ref{sec:numerics} is devoted to a full 
numerical simulation of the evolution under the model Hamiltonian, validating
and clarifying the analytical predictions. A brief discussion of the results is
found in Sec.~\ref{sec:conclusions}.

\section{Imperfect Quantum State Transfer}
\label{sec:IST}

The time evolution of a quantum state $|\phi(t)\rangle$ under the action of a
governing Hamiltonian $H$ is given by
\begin{equation}
|\phi(t)\rangle=e^{-itH/\hbar}|\phi(0)\rangle.
\label{eq:evol1}
\end{equation}
Suppose that $|\phi(t)\rangle$ is only defined at discrete sites
$j=1,2,\ldots,N$. Defining states $|i\rangle$ corresponding to unit basis
vectors associated with site $i$, one obtains the probability amplitudes 
$\langle i|\phi(t)\rangle=\phi_i(t)$. Given $N$ eigenvalues $\lambda^{(n)}$ and 
orthonormal eigenvectors $|\psi^{(n)}\rangle$ of $H$, Eq.~(\ref{eq:evol1}) 
becomes
\begin{equation}
|\phi(t)\rangle=\sum_{n=1}^{N}\exp\left[-\frac{it\lambda^{(n)}}{\hbar}\right]
|\psi^{(n)}\rangle\langle\psi^{(n)}|\phi(0)\rangle.
\label{eq:evol2}
\end{equation}
Suppose furthermore that the initial state is completely localized at a 
particular site $j$, so that $\langle i|\phi(0)\rangle=\delta_{ij}$.
Eq.~(\ref{eq:evol2}) can then be rewritten as
\begin{equation}
|\phi(t)\rangle=e^{-i\lambda^{(m)}t/\hbar}\sum_{n=1}^{N}e^{-it\left(
\lambda^{(n)}-\lambda^{(m)}\right)/\hbar}|\psi^{(n)}\rangle
\langle\psi^{(n)}|j\rangle.
\label{eq:evol3}
\end{equation}
Evidently if $(\lambda^{(n)}-\lambda^{(m)})t/\hbar=2\pi s$ ($s\in\mathbb{Z}$) 
for any $n$ and $m$, then $|\phi(t)\rangle=|j\rangle$ for all $j$ up to an
unimportant overall phase; that is, the discrete Hamiltonian is site periodic. 
This immediately implies that the Hamiltonian is site periodic if all the 
eigenvalues of $H$ satisfy the ratio condition
\begin{equation}
\frac{\lambda^{(n)}-\lambda^{(m)}}
{\lambda^{(p)}-\lambda^{(q)}}\in \mathbb{Q} 
,\label{eq:ratio}
\end{equation}
for all possible indices $\{n,m,p,q\}$ (except $\lambda^{(p)}=\lambda^{(q)}$). 
For large systems in practice, condition~(\ref{eq:ratio}) can only be satisfied
if the spectrum is linear, i.e.\ the gap between successive non-degenerate
eigenvalues is constant.

In perfect quantum state transfer (PST) between sites $i$ and $j$, a state 
initially localized in state $|i\rangle$ ends up in state $|j\rangle$, 
$j\neq i$ (and vice versa) after some elapsed time $t$; these sites are 
therefore periodic in time $2t$. PST therefore implies site periodicity, but
the converse is not generally true. For space-symmetric Hamiltonians that 
commute with the parity operator, however, site periodicity implies PST. 
Suppose that the smallest difference in eigenvalues is 
$\Delta=\lambda^{(n)}-\lambda^{(m)}$ for some $m$ and $n$; then site
periodicity occurs in time $t_P=2\pi\hbar/\Delta$. Consider an initial state 
in a superposition state of sites $i$ and $j$ equidistant from an axis of 
symmetry, $|\phi(0)\rangle=\frac{1}{\sqrt{2}}\left(|i\rangle+|j\rangle\right)$.
Because $|\phi(0)\rangle$ has even parity, only even-parity eigenvectors 
will contribute to the sum in Eq.~(\ref{eq:evol3}). Site periodicity must be 
maintained, but the absence of odd eigenvalues implies that the gap has now 
doubled to $\Delta'=2\Delta$ (assuming the spectrum is linear); now 
$t_P=4\pi\hbar/\Delta'$. At half this time $t=t_P/2=2\pi\hbar/\Delta'$ the sum 
over $n$ in Eq.~(\ref{eq:evol3}) still resolves to the identity. Thus, 
even-parity states such as the superposition state 
$\frac{1}{\sqrt{2}}\left(|i\rangle+|j\rangle\right)$ evolve to themselves in 
half the site periodicity time $t=\pi\hbar/\Delta$, which is only possible if
after this time $|i\rangle\leftrightarrow|j\rangle$, i.e.\ each site undergoes
PST to its mirror-symmetric counterpart. To summarize: mirror-symmetry and a
linear spectrum are together sufficient to ensure that the Hamiltonian supports
PST.

Unfortunately, most Hamiltonians do not possess a perfectly linear spectrum,
even if they are mirror symmetric. One may instead probe for IST, where the 
initial state at some site transfers to a different site with high probability.
A na\"\i ve approach is to approximate 
all eigenvalues in the spectrum by rationals with the same common denominator, 
so that in principle the ratio condition~(\ref{eq:ratio}) is automatically 
satisfied. Then in principle there should exist some time at which the 
Hamiltonian is almost site periodic, where initial states localized at a given 
site return with probability proportional to the accuracy of the rational 
approximation. By extension, if parity is a commuting operator then at half 
this time the probability at the mirror-symmetric vertex (the fidelity) should
approach unity to within a similar error.

In fact the criteria for IST are slightly more involved than the na\"\i ve
approach discussed above~\cite{Vinet2012b}. Recall that for a site-periodic
Hamiltonian there exists a time $\lambda^{(n)}t/\hbar=2\pi M_n$, 
$M_n\in\mathbb{Z}$ for all $n$ so that the sum over $n$ in Eq.~(\ref{eq:evol2}) 
resolves to the identity. For almost site periodicity, and allowing for an
arbitrary $n$-independent phase, this criterion would become 
$\left|\lambda^{(n)}t_P/\hbar+\varphi-2\pi M_n\right|<\delta$, where 
$\delta\ll 1$. 
For reflection-symmetric Hamiltonians the eigenvalues of even and odd parity 
eigenstates interleave. At the IST time $t=t_P/2$, odd-$n$ eigenstates map 
$|i\rangle$ to $|N-i+1\rangle$ but with an additional $\pi$ phase. The IST
condition then reads
\begin{equation}
-\delta<\frac{\lambda^{(n)}t}{\hbar}-\alpha_n-2\pi M_n<\delta,
\quad n=1,2,\dots,N,
\label{eq:inequality}
\end{equation}
where $\alpha_n=\pi n-\varphi$. The first task would be to find integers $M_n$, 
the phase $\varphi$, and time $t$ to satisfy this set of inequalities for each 
$n$ at fixed $\delta$. Once this is accomplished, one would calculate the 
output fidelity. Given initial occupation of vertex $1$, the desired output 
vertex is $N$. The output fidelity is therefore
\begin{equation}
F(t)=|\langle N|e^{-iHt}|1\rangle|^2.
\label{eq:fidelity}
\end{equation}
IST is said to occur at time $t$ if $F(t)$ exceeds some minimum threshold, for 
example $F_{\rm min}=0.9$.

\section{Model and behavior}
\label{sec:model}

\begin{figure}[t]
\centering
\includegraphics [width=0.48\textwidth, angle=0]{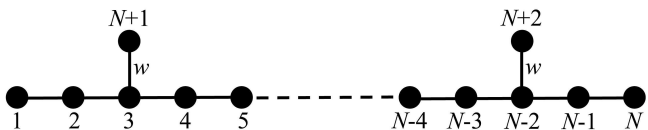}
\caption{Geometry of the model. Chain sites are labeled 1 through $N$, and
the tunneling amplitude between adjacent sites is constant $J$ (not labeled in
the figure). The impurities consist of additional leaves at sites 3 and $N-2$
labeled by $N+1$ and $N+2$, respectively. The amplitude to tunnel between chain 
and impurity sites is $w$.}
\label{fig:geometry}
\end{figure}

Consider an array of spin-$\frac{1}{2}$ particles, each confined to its own 
lattice site. As shown in Fig.~\ref{fig:geometry}, $N$ sites are arranged in a
one-dimensional chain with an additional spin connected to the third site from
each end. Given a spin-spin coupling constant $J$ along the chain (not 
explicitly labeled in the figure) and coupling constant $w$ between the 
additional spins and their counterparts on the chain, the XY Hamiltonian reads
\begin{equation}
H=H_{\rm 1D}+H',
\label{eq:spinHam}
\end{equation}
where 
\begin{equation} 
H_{\rm 1D}=\frac{J}{2}\sum_{i=1}^{N-1}\left(X_iX_{i+1}+Y_iY_{i+1}\right)
\end{equation}
corresponds to the one-dimensional uniform chain Hamiltonian and the 
Hamiltonian for the two additional impurity sites is
\begin{eqnarray} 
H'&=&\frac{w}{2}\big(X_3X_{N+1}+Y_3Y_{N+1}\nonumber \\
&+&X_{N-2}X_{N+2}+Y_{N-2}Y_{N+2}\big);
\end{eqnarray}
here $X=\sigma_x$ and $Y=\sigma_y$ are two-dimensional Pauli matrices. The 
total spin projection is a good quantum number and the Hamiltonian diagonalizes 
into blocks with a fixed number of excitations. Because only one excitation is
required in order to effect state transfer, it is conventional to work in the
single-excitation subspace~\cite{Christandl2005,Casaccino2009}. In this case, 
the spin Hamiltonian~(\ref{eq:spinHam}) is equivalent to a single particle 
hopping via a tight-binding Hamiltonian on an array with the same geometry, 
but with spin-coupling constants replaced by hopping 
amplitudes~\cite{Hines2007}:
\begin{equation} 
H_{\rm 1D}=J\sum_{i=1}^{N-1}\left(|i\rangle\langle i+1|+|i+1\rangle\langle i|
\right);
\label{eq:H1D}
\end{equation}
\begin{equation} 
H'=w\left(|3\rangle\langle N+1|+|N-2\rangle\langle N+2|+\mbox{H.c.}\right).
\label{eq:Hprime}
\end{equation}
Here, states $|i\rangle$ are unit vectors associated with site $i$, and H.c.\
stands for the Hermitian conjugate.

Determination of the conditions under which IST can occur (if any) for the
geometry shown in Fig.~\ref{fig:geometry} hinges on the diagonalization of the 
spin Hamiltonian~(\ref{eq:spinHam}) or alternatively (and more simply) the 
equivalent hopping Hamiltonian, Eqs.~(\ref{eq:H1D}) and (\ref{eq:Hprime}).
To probe numerically for IST, 
we explicitly obtain the time-dependence of the probability on the output 
vertex using Eq.~(\ref{eq:evol2}), and calculate the time-dependent fidelity 
$F(t)$, Eq.~(\ref{eq:fidelity}).

\begin{figure}[t]
\centering
\includegraphics[width=0.49\textwidth, angle=0]{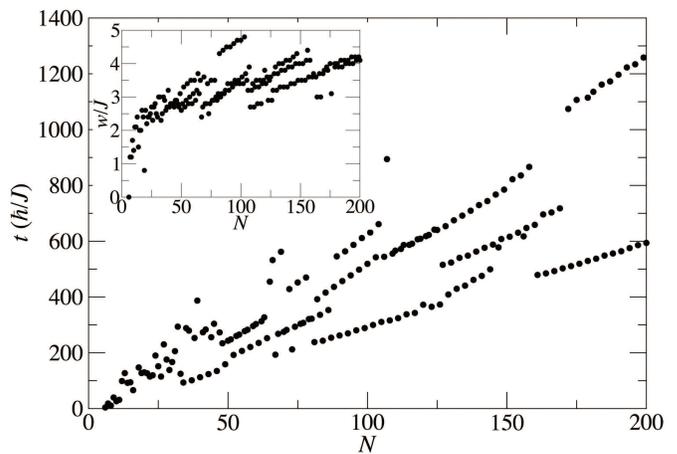}
\caption{The minimum time, in units of $\hbar/J$, for IST on the 
impurity-modified chain is shown as a function of the chain length for $N$ 
between 6 and 200 sites. The inset shows the value of the impurity hopping 
parameter (impurity spin coupling constant) $w$ associated with the minimum 
time, in units of the chain parameter $J$.}
\label{fig:alldata}
\end{figure}

Figure~\ref{fig:alldata} shows the minimum time for which IST is possible as a
function of chain length $N$, for $6\leq N\leq 200$. The results were obtained
by looping over values of $\tilde{w}=w/J$ in the range $0\leq\tilde{w}\leq 5$ 
in $0.1$ increments for each value of $N$. For each $\tilde{w}$, the value of 
$tJ/\hbar$ was increased in $0.1$ increments until $F(t)$ was found to exceed 
$F_{\rm min}=0.9$. Though IST is found to occur for many choices of $t$, only 
the lowest 
value of $t$ for each $\tilde{w}$ is shown in Fig.~\ref{fig:alldata}; the value 
of $\tilde{w}$ that minimizes $t$ is shown in the inset. The data clearly show 
that the minimum time (and the impurity coupling constants associated with 
these) are nonmonotonic in $N$. There are intervals where the minimum time 
appears to scale linearly with $N$, but these are interrupted and interspersed 
with different trends. Likewise, the impurity coupling constants seem to scale 
roughly as $\sqrt{N}$, but the data are not clean. 

In principle, one could try to determine the conditions on $\tilde{w}$ that 
yield the absolute lowest-slope curve. This would yield a slightly modified
spin chain where the IST time would scale linearly with length, albeit with
restrictions on the values of $N$ for which IST is possible. In this work, 
however, we pursue a different tack; namely, determining values of $w$ valid 
for any instance of $N$ and for which the IST time scales efficiently with $N$ 
(i.e.\ as a power-law with $N$ preferably with a low exponent). The analytical 
treatment discussed in the next section addresses this strategy.

\section{Analytical treatment}
\label{sec:analytics}
 
While a complete analytical solution for arbitrary $w$ (or for arbitrary $w/J$ 
choosing $J$ as the characteristic energy scale) appears difficult to obtain, 
approximate solutions may be obtained by solving the problem in the vicinity of 
the additional site(s) and matching to the bulk solution. First consider the 
left block in Fig.~\ref{fig:geometry}, consisting only of sites 
$i=\{1,2,3,4,5,N+1\}$ (the block with sites $i\to N-i+1$, $i=1,2,\ldots,N$, and 
$N+1\to N+2$ is wholly equivalent). 
Expressing the left block solution as 
\begin{equation}
|\psi_{\rm L}\rangle=\sum_{i=1}^4a_i|i\rangle+a_{N+1}|N+1\rangle,
\end{equation}
and operating with the left-block hopping Hamiltonian
\begin{eqnarray} 
H_{\rm L}&=&J\sum_{i=1}^4\left(|i\rangle\langle i+1|+|i+1\rangle\langle i|
\right)
\nonumber \\
&+&w\left(|3\rangle\langle N+1|+|N+1\rangle\langle 3|\right)
\label{eq:hopHamL}
\end{eqnarray}
such that $H_{\rm L}|\psi_{\rm L}\rangle=\lambda|\psi_{\rm L}\rangle$, one 
obtains
\begin{eqnarray}
&&\{a_1,a_2,a_3,a_4,a_{N+1}\}\nonumber \\
&&\quad=\frac{a_5}{\gamma}\left\{1,\tilde{\lambda},\tilde{\lambda}^2-1,
\frac{\tilde{\lambda}^2-1+\gamma}{\tilde{\lambda}},
\tilde{w}\frac{\tilde{\lambda}^2-1}{\tilde{\lambda}}\right\},
\label{eq:leftsol}
\end{eqnarray}
where $\gamma=1+\tilde{w}^2-(3+\tilde{w}^2)\tilde{\lambda}^2
+\tilde{\lambda}^4$, and $\tilde{w}=w/J$ and $\tilde{\lambda}=\lambda/J$ are 
the rescaled values of $w$ and $\lambda$ relative to the characteristic energy 
scale $J$.

The most important eigenvectors for 
IST are those with large amplitudes on the first and last sites of the chain, 
and by extension small amplitudes elsewhere. Small amplitude on the third site 
can be ensured if $\tilde{\lambda}\sim 1$. Likewise, for the amplitude on the 
fourth site to be small one requires 
$\gamma/\tilde{\lambda}\approx\gamma\ll 1$; this 
is possible if one chooses $\tilde{w}\gg 1$. More concretely, suppose one sets
$a_1=\alpha a_4$ assuming $\alpha\gg 1$; using the explicit coefficients in 
Eq.~(\ref{eq:leftsol}) one obtains
\begin{equation}
\tilde{w}=\sqrt{\tilde{\lambda}_c(\alpha\tilde{\lambda}_c^3-2\alpha
\tilde{\lambda}_c-1)\over\alpha(\tilde{\lambda}_c^2-1)}.
\end{equation}
Setting $\tilde{\lambda}_c=1-\epsilon$ and expanding to lowest order in 
$\epsilon\sim 0$ gives
\begin{equation}
\tilde{w}\approx\pm\sqrt{1+\alpha\over 2\alpha\epsilon}
\approx\pm\frac{1}{\sqrt{2\epsilon}}.
\label{eq:wapprox}
\end{equation}
Note that necessarily $\epsilon>0$ which ensures that 
$\tilde{\lambda}_c\lesssim 1$. Thus, the eigenvalue that ensures that the
amplitude on the endpoint(s) is resonantly enhanced is
\begin{equation}
\tilde{\lambda}_c\approx 1-\epsilon\approx 1-\frac{1}{2\tilde{w}^2}.
\label{eq:lambdaapprox}
\end{equation}
The value of $\tilde{w}$ 
can be positive or negative. Inserting this expression for $\tilde{\lambda}$ 
into Eq.~(\ref{eq:leftsol}) and again assuming $\tilde{w}\gg 1$ gives
\begin{equation}
\{a_1,a_2,a_3,a_4,a_{N+1}\}\propto
\left\{\tilde{w}^2,\tilde{w}^2,-1,-\frac{1}{4},-\tilde{w}\right\}.
\label{eq:leftsolbigw}
\end{equation}
Note that the amplitudes on both the first and second sites of the chain are 
much larger than those elsewhere, which implies that any possible IST could 
equally originate at either of these sites. Thus, $\tilde{\lambda}_c$ is the 
eigenvalue for a state strongly localized in the vicinity of the impurity.

Next consider the solutions $|\psi_{\rm 1D}\rangle$ for the bulk 
one-dimensional chain $H_{\rm 1D}$, Eq.~(\ref{eq:H1D}):
\begin{equation}
\langle i|\psi_{\rm 1D}^{(n)}\rangle=\sin(k_ni);\;
\tilde{\lambda}_{\rm 1D}^{(n)}=2\cos(k_n).
\label{eq:bulksol}
\end{equation}
The reflection symmetry imposes the constraint that 
$\langle i|\psi_{\rm 1D}^{(n)}\rangle
=\pm\langle N-i+1|\psi_{\rm 1D}^{(n)}\rangle$, i.e.\ that the solution is an
eigenstate of the parity operator. Setting 
$\sin(k_ni)=-\sin(k_ni)\cos[k_n(N+1)]+\cos(k_ni)\sin[k_n(N+1)]$ for the 
even-parity case gives the conditions $\cos[k_n(N+1)]=-1$ and 
$\sin[k_n(N+1)]=0$ which is satisfied by $k_n=\frac{\pi(2n+1)}{N+1}$, 
$n=1,2,\ldots$. Likewise, the odd-parity solution requires 
$k_n=\frac{2\pi n}{N+1}$, so that overall $k_n=\frac{\pi n}{N+1}$, 
$n=0,1,\ldots,N-1$. The invariance of the probability density under parity 
implies that all eigenvalues appear in plus/minus pairs, as 
$2\cos(k_n)\to-2\cos(k_n)$ for $n\to N+1-n$ ($k_n\to\pi-k_n$).


The bulk eigenvectors should automatically match the left-block 
solution~(\ref{eq:leftsol}) when $\tilde{w}=0$. Whether using the bulk solution 
$\sin\left[\cos^{-1}\left(\tilde{\lambda}/2\right)i\right]$ from 
Eq.~(\ref{eq:bulksol}) or setting $\tilde{w}=0$ in Eq.~(\ref{eq:leftsol}), one
obtains the same result:
\begin{equation}
\{a_1,a_2,a_3,a_4\}\propto\left\{1,\tilde{\lambda},
\tilde{\lambda}^2-1,\tilde{\lambda}\left(\tilde{\lambda}^2-2\right)\right\}.
\label{eq:solwzero}
\end{equation}
Alternatively, one can require that the bulk solution matches the amplitude on
the last site of the left block:
\begin{equation}
\sin(4k_n)=\frac{\sin(5k_n)(\tilde{\lambda}^2-1+\gamma)}{\gamma\tilde{\lambda}}
=\frac{\sin(5k_n)\tilde{\lambda}(\tilde{\lambda}^2-2)}
{\tilde{\lambda}^4-3\tilde{\lambda}^2+1}.
\end{equation}
Using trigonometric identities it is straightforward to verify that this 
condition is (non-uniquely) satisfied by choosing $\tilde{\lambda}
=\tilde{\lambda}^{(n)}=2\cos(k_n)$, as expected for the one-dimensional chain.

Return again to the $\tilde{w}\gg 1$ case, but now for generic 
$\tilde{\lambda}$ (keeping in mind that the $\tilde{\lambda}\sim 1$ are of 
particular interest, and that the reflection symmetry of the site array implies 
that all eigenvalues have negative counterparts 
$\tilde{\lambda}\to -\tilde{\lambda}$). The parameter $\gamma$ in 
Eq.~(\ref{eq:leftsol}) becomes 
$\gamma\to\tilde{w}^2\left(1-\tilde{\lambda}^2\right)$. The left-block 
solution for $\tilde{w}\gg 1$ therefore becomes
\begin{eqnarray}
&&\{a_1,a_2,a_3,a_4,a_{N+1}\}\to\Bigg\{1,\tilde{\lambda},\tilde{\lambda}^2-1,
\nonumber \\
&&\qquad\frac{\tilde{w}^2(1-\tilde{\lambda}^2)
+\tilde{\lambda}^4-3\tilde{\lambda}^2+1}{\tilde{\lambda}},
\frac{\tilde{w}(\tilde{\lambda}^2-1)}{\tilde{\lambda}}\Bigg\},\hphantom{aaa}
\label{eq:leftsolbigw2}
\end{eqnarray}
neglecting unimportant prefactors.
Comparison of Eqs.~(\ref{eq:solwzero}) and (\ref{eq:leftsolbigw2}) reveals that
the $\tilde{w}=0$ and $\tilde{w}\gg 1$ wavefunctions match at all but the 
fourth site in the chain: if $\tilde{\lambda}\sim 1$ then $a_{N+1}\to 0$. This 
implies that the $\tilde{w}\gg 1$ 
eigenvalues almost coincide with those of the impurity-free chain. On the face 
of it, this is disappointing, as the bare chain cannot support efficient IST. 
Yet Eq.~(\ref{eq:lambdaapprox}) clearly states that there always exists one 
eigenvalue for the localized state near the impurity that varies as 
$1-1/2\tilde{w}^2$, approaching unity asymptotically. The implication is that 
there must be (avoided) level crossings at finite $\tilde{w}$ for every 
$\tilde{\lambda}^{(n)}$ that comes into resonance with the eigenvalue 
$\tilde{\lambda}_c$ of the localized state. It turns out that efficient IST 
hinges on the first of these avoided crossings.

Target eigenvalues are those in the vicinity of (but just below) unity, 
$\tilde{\lambda}^{(n)}=2\cos(k_n)-\epsilon\sim 1^-$, so that $k_n\gtrsim\pi/3$.
That said, the $k_n=\pi n/(N+1)$ are discrete even in the limit of large $N$. 
There are therefore three different cases to consider: $N=3m$ and $3m+1$
with $m=2,3,\ldots$, and $3m-1$ with $m=3,4,\ldots$. Consider first the $N=3m$ 
case. The wave vectors of interest are indexed by $n=m+r$:
\begin{equation}
k_{m+r}=\frac{\pi(m+r)}{3m+1},
\end{equation}
where $r=1,2,\ldots$ for $k_{m+r}>\pi/3$ to ensure that the associated 
eigenvalues $\tilde{\lambda}_r<1$ (subscripts are now used to remind the reader 
that $r=1$ corresponds to $n=3m+r$). The $r=1$ eigenvalue 
$\tilde{\lambda}_1\approx 2\cos(k_{m+1})$ must cross the critical eigenvalue
$\tilde{\lambda}_c$ in Eq.~(\ref{eq:lambdaapprox}) at some value of 
$\tilde{w}\gg 1$. The behavior 
of the system for large chains is of particular interest; expanding around 
$N=3m\gg 1$ and large $\tilde{w}$, the levels cross when 
\begin{equation}
1-\frac{2\pi}{\sqrt{3}N}\approx 1-\frac{1}{2\tilde{w}^2},
\end{equation}
which yields the critical impurity coupling constant
\begin{equation}
\tilde{w}_c\approx\pm\frac{3^{1/4}}{2\sqrt{\pi}}\sqrt{N}\approx\pm 0.37\sqrt{N}.
\label{eq:weight0}
\end{equation}

Alternatively, consider the left-block eigenvector~(\ref{eq:leftsol}), which 
becomes
\begin{eqnarray}
&&\{a_1,a_2,a_3,a_4,a_{N+1}\}\approx\Bigg\{\frac{\sqrt{3}}{2}
+\frac{8\pi\tilde{w}^2}{N}+\mathcal{O}\left(\frac{\tilde{w}^4}{N^2}\right),
\nonumber \\
&&\frac{\sqrt{3}}{2}+\frac{8\pi\tilde{w}^2}{N}
+\mathcal{O}\left(\frac{\tilde{w}^4}{N^2}\right),-\frac{8\pi}{N}
+\mathcal{O}\left(\frac{\tilde{w}^2}{N^2}\right),\nonumber \\
&&-\frac{\sqrt{3}}{2}-\frac{16\pi}{3N}
+\mathcal{O}\left(\frac{\tilde{w}^2}{N^2}\right),
-\frac{8\pi\tilde{w}}{N}+\mathcal{O}\left(\frac{\tilde{w}^3}{N^2}\right)\Bigg\}.
\end{eqnarray}
Clearly, the expansions above are analytic only if $\tilde{w}$ varies with $N$
more slowly than $\sqrt{N}$ (or one could obtain a convergent series by 
expanding the solution~(\ref{eq:leftsol}) in $N/\tilde{w}$ for $\tilde{w}$ a 
polynomial in $N$ with exponent greater than $1/2$). The critical case 
$\tilde{w}_c=\alpha\sqrt{N}$ is therefore of particular interest. For large 
$N$ one obtains
\begin{eqnarray}
&&\{a_1,a_2,a_3,a_4,a_{N+1}\}\approx\Bigg\{
\frac{\sqrt{3}}{2\left(1-\frac{4\pi\alpha^2}{\sqrt{3}}\right)},
\frac{\sqrt{3}}{2\left(1-\frac{4\pi\alpha^2}{\sqrt{3}}\right)},
\nonumber \\
&&-\frac{2\pi}{N\left(1-\frac{4\pi\alpha^2}{\sqrt{3}}\right)},
-\frac{\sqrt{3}}{2},
-\frac{2\pi\alpha}{\sqrt{N}\left(1-\frac{4\pi\alpha^2}{\sqrt{3}}\right)}\Bigg\},
\label{eq:resonance}
\end{eqnarray}
which matches the bulk solution at the fourth site. The amplitudes on the first
and second sites are strongly enhanced relative to the others if one sets
$1-\frac{4\pi\alpha^2}{\sqrt{3}}=0$ or $\alpha=3^{1/4}/2\sqrt{\pi}$, consistent
with Eq.~(\ref{eq:weight0}). Equivalently, the state~(\ref{eq:resonance}) has 
maximal overlap with the state~(\ref{eq:leftsolbigw}) when the first two 
amplitudes above equal $\tilde{w}^2=\alpha^2N$, which also occurs for 
$\alpha=3^{1/4}/2\sqrt{\pi}$. 

The $N=3m+1$ and $N=3m-1$ cases proceed analogously. For $N=3m+1$ for large 
$N$ one obtains the critical impurity coupling constant
\begin{equation}
\tilde{w}_c\approx\pm\frac{3^{1/4}}{\sqrt{2\pi}}\sqrt{N}\approx 0.53\sqrt{N},
\label{eq:weight1}
\end{equation}
while for $N=3m-1$ in the same limit one obtains
\begin{equation}
\tilde{w}_c\approx\frac{1}{3^{1/4}\sqrt{2\pi}}\approx 0.30\sqrt{N}.
\label{eq:weight2}
\end{equation}
While all critical impurity coupling constants scale as $\sqrt{N}$, the 
prefactor depends on the particular choice of $N$~(mod 3), $N+1$~(mod 3), or 
$N-1$~(mod 3). 

Return again to the $N=3m$ case. As $\tilde{w}$ is increased through the
critical value~(\ref{eq:weight0}), the $2\cos(k_{m+1})$ eigenvalue must exhibit 
and an avoided crossing with $\tilde{\lambda}_c$, while its associated 
eigenvector strongly mixes with the next odd-parity state with eigenvalue 
$\tilde{\lambda}_3$. Likewise, the eigenvalue 
$\tilde{\lambda}_2$ of the first relevant even-parity state should also follow 
Eq.~(\ref{eq:lambdaapprox}) for large $\tilde{w}$, while strongly mixing with 
the $\tilde{\lambda}_4$ state for intermediate $\tilde{w}$ at the second 
avoided crossing in the vicinity of $2\cos(k_{m+3})$, etc. That said, 
presumably the $\tilde{\lambda}_2$ state only mixes weakly with the 
$\tilde{\lambda}_4$ state near the first avoided crossing (and of course not at
all with the $\tilde{\lambda}_1$ and $\tilde{\lambda}_3$ states due to parity),
which should occur for much larger values of $\tilde{w}$ than the second 
avoided crossing. With this assumption, $\tilde{\lambda}_2$ follows 
Eq.~(\ref{eq:lambdaapprox}) throughout the first level crossing. 

By inference, therefore, only three states are relevant to the first avoided 
crossing, corresponding to eigenvalues indexed by $r=1,2,3$. The same 
phenomenon also applies to the $N=3m+1$ and $N=3m-1$ cases, but the avoided 
crossing occurs for different $k$ labels and therefore at different energies.
Importantly, all three states involved in the avoided crossing at $\tilde{w}_c$
have strongly enhanced amplitude on the first and last site of the chain (as
well as the second and second-from last), of order unity after normalization. 
All other states will be far off-resonant, and will have low amplitudes on the 
endpoints proportional to the overall normalization constant for bulk 
eigenvectors $\propto\sqrt{2/N}$. But the sum of outer products of eigenvectors 
must resolve to the identity. This implies that as $N\to\infty$, no 
off-resonant eigenvectors will contribute to the state transfer: the sum in 
Eq.~(\ref{eq:evol2}) will only include resonant eigenvectors. Because only
three equally-spaced eigenvalues are involved (plus their negatives), the state 
transfer must be asymptotically perfect. 

Obtaining an analytical estimate of the energy splitting at the critical 
impurity coupling constant is not as straightforward as it appears. The usual 
method would be to start with eigenfunctions $|\psi^{(m+1)}_{\rm 1D}\rangle$ 
and 
$|\psi^{(m+3)}_{\rm 1D}\rangle$ of the unperturbed Hamiltonian~(\ref{eq:H1D}) 
and then calculate the mixing caused by the perturbation~(\ref{eq:Hprime}), 
i.e.\ the off-diagonal term of the mixing matrix
$\Delta\equiv\langle\psi^{(m+1)}_{\rm 1D}|H'|\psi^{(m+3)}_{\rm 1D}\rangle$. 
The impurities have no support on the bare chain, however, so in principle the
energy splitting $\Delta=0$. One can nevertheless estimate $\Delta$ as 
follows. The contribution to $\Delta$ from all the chain sites will be zero,
as the unperturbed eigenfunctions are orthogonal. At $\tilde{w}_c$, the 
amplitude on the impurity site for the $k=m+1$ state is $-1/2\tilde{w}_c
=-\sqrt{\pi}/3^{1/4}\sqrt{N}$ after normalization (which is dominated by the
amplitudes on the first two and last two sites of the chain). Likewise, the 
amplitude on the impurity site for the $k=m+3$ state is 
$-\sqrt{2/N}(4\sqrt{\pi}/3^{3/4}\sqrt{N})=-4\sqrt{2\pi}/3^{3/4}N$ including the
normalization factor for $\sin(k_n)$ eigenfunctions. Because the action of the 
Hamiltonian on this site returns the same amplitude (the energy is almost 
unity), one obtains 
\begin{equation}
\Delta\approx\frac{4\sqrt{2\pi}}{3^{3/4}N}
\frac{\sqrt{\pi}}{3^{1/4}\sqrt{N}}=\frac{4\sqrt{2}\pi}{3N^{3/2}}.
\label{eq:splitting}
\end{equation}
While the coefficient is probably not that accurate, the analytics suggest that
the energy splitting at the avoided crossing scales with the chain length as 
$N^{-3/2}$. The IST time should therefore scale as $t_{\rm IST}\sim N^{3/2}$.

At this juncture the reader might well be wondering what is special about 
adding impurities to the third and the third-from-last sites of the chain. 
Suppose that the impurities were instead located on the second and 
second-from-last sites of the chain. The left-block state, analogous to 
Eq.~(\ref{eq:leftsol}) is found to be
\begin{equation}
\{a_1,a_2,a_3,a_{N+1}\}
=\frac{a_4}{\gamma\tilde{\lambda}}\left\{1,\tilde{\lambda},\tilde{\lambda}^2-1
-\tilde{w}^2,
\tilde{w}\right\},
\label{eq:leftsol2}
\end{equation}
where $\gamma=2-\tilde{\lambda}^2-\tilde{w}^2$. The amplitude on the first site 
can be made larger than in the bulk only if $\tilde{\lambda}\sim 1$ and 
$\tilde{w}\sim 0$. The
second condition is unfortunately equivalent to the unmodified chain, and is
therefore not useful. Consider instead impurities on the fourth and 
fourth-from-last sites of the chain. The left-block solution is now
\begin{eqnarray}
&&\{a_1,a_2,a_3,a_4,a_5,a_{N+1}\}\nonumber \\
&&=\frac{a_6\tilde{\lambda}}{\gamma}\left\{1,\tilde{\lambda},
\tilde{\lambda}^2-1,\tilde{\lambda}(\tilde{\lambda}^2-2),
\tilde{\lambda}^2-2+\gamma,\tilde{w}(\tilde{\lambda}^2-2)\right\},
\nonumber \\
\label{eq:leftsol4}
\end{eqnarray}
where $\gamma=3+2w^2-(4+w^2)\tilde{\lambda}^2+\tilde{\lambda}^4$. Following the 
analysis above, small amplitude on the fourth and fifth sites requires 
$\tilde{w}\gg 1$ and $\tilde{\lambda}\approx\sqrt{2}-1/2\sqrt{2}w^2$. To 
leading order in $\tilde{w}$ one obtains
\begin{eqnarray}
&&\{a_1,a_2,a_3,a_4,a_5,a_{N+1}\}\nonumber \\
&&\approx\left\{\tilde{w}^2,\sqrt{2}\tilde{w}^2,\tilde{w}^2,-\sqrt{2},
-\frac{9}{8},-\tilde{w}\right\}.
\label{eq:leftsol4b}
\end{eqnarray}
Just as for impurities on sites $i=3$ and $N-i+2$, there is a strong 
enhancement of amplitude on the first and last sites. This enhancement is now 
shared with four other sites ($i=2$, $3$, $N-1$, and $N-2$), however, 
decreasing the total amplitude available on the target sites. This trend 
continues as the impurities move further from the ends of the chain. Thus, 
adding impurities to the third and third-from-last sites is optimal.

\section{Numerical Results}
\label{sec:numerics}

The strategy pursued in this work is to determine if there exist 
values of $\tilde{w}(N)$ that allow APST to occur for all $N$, but perhaps not 
at the absolute minimum time allowable. Using the analytical results of the
previous section as a guide, for particular values of $\tilde{w}$ one expects
that for eigenvalues $\tilde{\lambda}^{(n)}\lesssim 1$ of the Hamiltonian the 
associated eigenvectors $|\psi^{(n)}\rangle$ will have a large overlap with the 
first and last sites of the chain. If the eigenvalues are reverse ordered so
that $\tilde{\lambda}^{(n)}\geq\tilde{\lambda}^{(n+1)}$ for $n=1,2,\ldots,N-1$,
then there exist values of $n\gtrsim N/3$ where 
$\tilde{\lambda}^{(n)}\lesssim 1$. Define $N\equiv 3m+p$, where 
$p\in\{-1,0,1\}$. Then one can 
define $\tilde{\lambda}_q=\tilde{\lambda}^{(q+m)}$ which are all less than
unity for $q=1,2,\ldots$, and their associated eigenvectors $|\psi_q\rangle$. 
Then IST should result for values of $\tilde{w}$ where the eigenvalues and 
eigenvectors satisfy the following two criteria:
\begin{eqnarray}
\left|\frac{\Delta_2-\Delta_1}{\Delta_1}\right|&\leq&\epsilon;
\label{eq:gapdiff}\\
2\sum_{q=1}^3\left|\langle\psi_q|1\rangle\right|^2&\geq&f,
\label{eq:overlaps}
\end{eqnarray}
where the two successive eigenvalue gaps are 
$\tilde{\Delta}_1=\tilde{\lambda}_1-\tilde{\lambda}_2$ and 
$\tilde{\Delta}_2=\tilde{\lambda}_2-\tilde{\lambda}_3$, and $\epsilon\ll 1$ and 
$f\lesssim 1$ are free parameters. 

\begin{figure}[t]
\centering
\includegraphics[width=0.49\textwidth, angle=0]{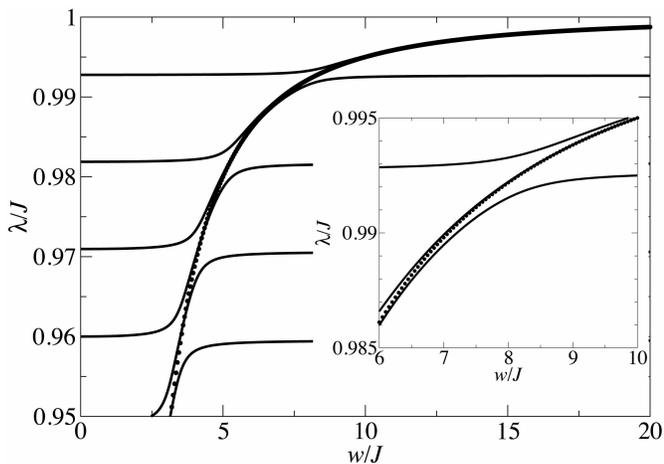}
\caption{The first few eigenvalues 
$\tilde{\lambda}_q=\lambda^{q+m}/J$ smaller
than unity ($q=1,2,\ldots,6$) are shown as a function of the impurity coupling 
parameter $\tilde{w}=w/J$ for chain length $N=3m=501$ (black lines). Also shown 
is the left-block eigenvalue $\tilde{\lambda}=\lambda/J=1-1/2\tilde{w}^2$ 
(dotted curve). The first avoided level crossing, closest to unity, is 
magnified in the inset.}
\label{fig:eigs}
\end{figure}

The Hamiltonian $H$ was diagonalized for each chain length $N$ in the range
$6\leq N\leq 501$. Figure~\ref{fig:eigs} shows the representative behavior of 
the first few eigenvalues less than unity as a function of the impurity 
coupling $\tilde{w}=w/J$ for $N=3m=501$. The salient features predicted by the
analytical treatment are readily observed here. The bare-chain eigenvalues 
$\tilde{\lambda}_q^0=2\cos[\pi(q+m)/(N+1)]$ (for $\tilde{w}=0$) are crossed by 
the eigenvalue $\tilde{\lambda}_c=1-1/2/\tilde{w}^2$ of the localized state at 
finite $w$, (almost) recovering their values for large $\tilde{w}$. The $q=1$ 
and $q=2$ eigenvalues both follow $\tilde{\lambda}_c$ for large $\tilde{w}$, 
while the $q=3$ eigenvalue approaches $2\cos[\pi(1+m)/(N+1)]$ in the same 
limit. 

The strong mixing between the localized and extended bulk eigenvectors gives 
rise to avoided crossings, and the first such crossing (closest to unity) is 
shown in the inset of Fig.~\ref{fig:eigs}. As expected, the $q=2$ eigenvalue 
closely follows $\tilde{\lambda}_c$ through the first avoided crossing, as it 
only weakly mixes with the $q=4$ eigenvalue here. At the value of $\tilde{w}$ 
where $\tilde{\lambda}_c\approx\tilde{\lambda}_2=\tilde{\lambda}_1^0$, the 
$\tilde{\lambda}_1$ and $\tilde{\lambda}_3$ eigenvalues are split equally above
and below $\tilde{\lambda}_2$. Thus, $\tilde{w}_c$ both defines the point at
which $\tilde{\lambda}_c=\tilde{\lambda}_1$ and the point at which the energy
gaps coincide, $\tilde{\Delta}_1=\tilde{\Delta}_2\equiv\tilde{\Delta}$.

Figure~\ref{fig:eigs} also suggests that there are multiple values of the 
impurity hopping parameter that give rise to eigenvalue resonances. The second 
avoided crossing occurs when $\tilde{\lambda}_c=\tilde{\lambda}_2$. For the
$N=501$ case displayed here, the analog of result~(\ref{eq:weight0}) for the 
second avoided crossing is $\tilde{w}_c'=(3^{1/4}/\sqrt{10\pi})\sqrt{N}
\approx 0.23\sqrt{N}$. While the qualitative behavior of the eigenvalue 
splitting is similar, one would expect a lower IST output fidelity as the 
maximum amplitude on the endpoint sites is now approximately $63\%$ lower than
for the first avoided crossing according to Eq.~(\ref{eq:leftsolbigw}).

\begin{figure}[t]
\centering
\includegraphics[width=0.49\textwidth, angle=0]{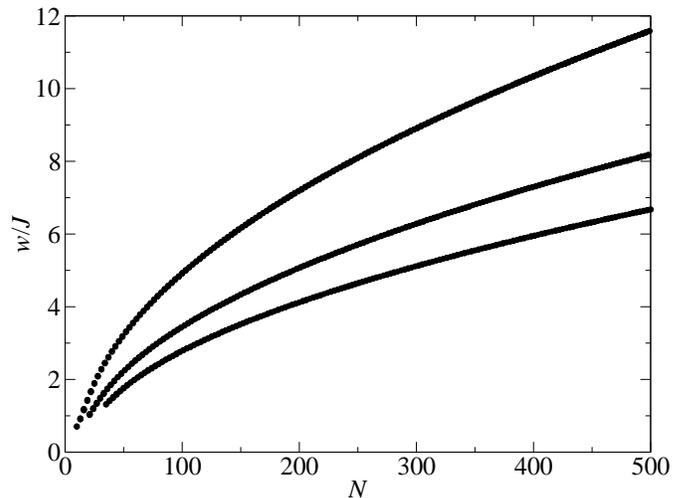}
\caption{The numerical value of the impurity coupling constant $\tilde{w}=w/J$ 
ensuring that the eigenvales at the first level crossing are equally spaced are 
shown as a function of the chain length $N$. Top, middle, and lower curves 
correspond to $N=3m+1$, $3m$, and $3m-1$ ($m$ integer), respectively. All 
curves closely follow $\tilde{w}\sim\sqrt{N}$ for large $N$.}
\label{fig:weightsnumeric}
\end{figure}

Eq.~(\ref{eq:gapdiff}) specifies that the spacing between successive
eigenvalues just less than unity be equal to within some tolerance $\epsilon$.
Figure~\ref{fig:weightsnumeric} shows the value of impurity parameter 
$\tilde{w}=w/J$ satisfying Eq.~(\ref{eq:gapdiff}) at the first avoided crossing
to a tolerance $\epsilon=10^{-10}$, for $6\leq N\leq 501$. There are three 
distinct curves, depending on the value of $N$. The lowest, middle, and upper 
curves correspond to $N+1$~(mod~3), $N$~(mod~3), and $N-1$~(mod~3), 
respectively. For the largest chain sizes considered, in the range 
$400\leq N\leq 500$, each of these curves is well fit (correlation coefficient 
$>0.999$) by a power law that closely resembles the analytical results, 
Eqs.~(\ref{eq:weight1}), (\ref{eq:weight0}), and (\ref{eq:weight2}), 
respectively: $\tilde{w}\approx 0.48N^{0.51}$ for $N=3m+1$, 
$\tilde{w}\approx 0.33N^{0.52}$ for $N=3m$, and $\tilde{w}\approx 0.27N^{0.52}$
for $N=3m-1$. In all cases, the results are compatible with a $\sqrt{N}$ 
scaling for large $N$. That said, the prefactors obtained numerically are all
found to be below the analytical preductions by approximately $10\%$; it is 
possible that the correspondence between analytics and numerics would tighten 
up for larger $N$.

\begin{figure}[t]
\centering
\includegraphics[width=0.49\textwidth, angle=0]{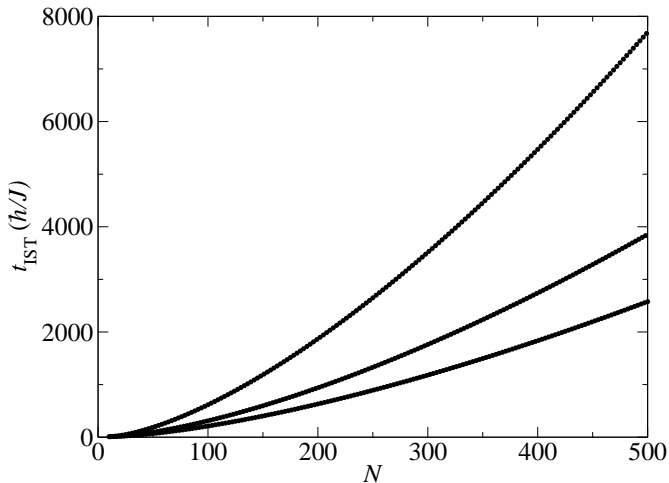}
\caption{The numerical estimate of the imperfect state transfer time 
$t_{\rm IST}$, based on the value of the resonant energy gaps, is shown as a 
function of the chain length $N$. Top, middle, and lower curves correspond to
$N=3m+1$, $3m$, and $3m-1$ ($m$ integer), respectively. All curves closely 
follow $t_{\rm IST}\sim N^{3/2}$ for large $N$.}
\label{fig:ISTtime}
\end{figure}

Under the assumption that only the first three eigenvectors $|\psi_q\rangle$ 
with eigenvalues $\lambda_q$ just below unity (and their negative counterparts) 
contribute to the transfer dynamics, the IST time can be predicted using 
Eq.~(\ref{eq:evol3}). Following the discussion in Sec.~\ref{sec:IST}, imperfect 
site periodicity occurs at a time $t=2\pi\hbar/(\lambda_1-\lambda_2)
=2\pi\hbar/\Delta_1$, and parity conservation implies that the imperfect state 
transfer time is half this, $t_{\rm IST}\equiv\pi\hbar/\Delta_1
=(\pi\hbar/J)/\tilde{\Delta}_1$. In rescaled time units $t=(\hbar/J)\tilde{t}$,
one obtains $\tilde{t}_{\rm IST}\equiv\pi/\tilde{\Delta}$, where 
$\tilde{\Delta}=\tilde{\Delta}_1=\tilde{\Delta}_2$. The state transfer times 
$t_{\rm IST}$ are shown as a function of the chain
length in the range $10\leq N\leq 501$ in Fig.~\ref{fig:ISTtime}. These times
are based on the energy gaps $\tilde{\Delta}_1=\tilde{\Delta}_2$ at the 
critical weight plotted in Fig.~\ref{fig:weightsnumeric}. As was the case for 
the critical weights, the IST time scales differently depending on the value of
$N$ (mod 3). For $N=3m$, the numerical results are best fit by the function
$\tilde{t}_{\rm IST}\approx 0.29N^{1.53}$ for large integer m; for $N=3m+1$ and 
$3m-1$ one obtains 
$\tilde{t}_{\rm IST}\approx 0.57N^{1.53}$ and 
$\tilde{t}_{\rm IST}\approx 0.20N^{1.52}$, respectively. The power laws are all 
consistent with a $N^{3/2}$ scaling of time, as predicted by the analytics 
discussed in the previous section. The prefactors also appear to scale roughly 
with those for the critical weights.

It is worthwhile to investigate the IST dynamics governed by 
Eq.~(\ref{eq:evol2}) more closely. Keeping in mind that $|\psi_1\rangle$ and 
$|\psi_3\rangle$ have even parity while $|\psi_2\rangle$ has odd parity (and
vice versa for $\lambda_q\to -\lambda_q$), so that for example 
$\langle\psi_1|1\rangle=\langle\psi_1|N\rangle$ while 
$\langle\psi_2|1\rangle=-\langle\psi_2|N\rangle$, one obtains
\begin{eqnarray}
\langle N|e^{-iHt/\hbar}|1\rangle&\approx&-2i\sin\left[\left(\lambda_2+\Delta
\right)t/\hbar\right]\left|\langle\psi_1|1\rangle\right|^2\nonumber \\
&+&2i\sin\left[\lambda_2t/\hbar\right]\left|\langle\psi_2|1\rangle\right|^2
\nonumber \\
&-&2i\sin\left[\left(\lambda_2-\Delta\right)t/\hbar\right]\left|
\langle\psi_3|1\rangle\right|^2,
\label{eq:ISTa}
\end{eqnarray}
keeping only resonant eigenvectors in the sum. Choosing $t$ such that the 
coefficients of each $\left|\langle\psi_q|1\rangle\right|^2$ term are equal, 
one obtains
$t_{\rm IST}=\pi\hbar/\Delta$, as expected. Then Eq.~(\ref{eq:ISTa}) becomes
\begin{eqnarray}
\langle N|e^{-iHt/\hbar}|1\rangle&=&2i\sin\left(\frac{\pi\lambda_2}{\Delta}
\right)\sum_{q=1}^3\left|\langle\psi_q|1\rangle\right|^2\nonumber \\
&\approx&i\sin\left(\frac{\pi\lambda_2}{\Delta}\right),
\label{eq:ISTb}
\end{eqnarray}
where the second line is obtained by assuming that only these eigenvectors 
(and their negative-eigenvalue counterparts) resolve the identity. 
Equation~(\ref{eq:ISTb}) appears to suggest that the maximum fidelity is
$F(t_{\rm IST})=\sin^2(\pi\lambda_2/\Delta)\approx\sin^2(\pi/\tilde{\Delta})$. 
The interference arises from the fact that while the first three eigenvalues 
below unity become approximately evenly spaced by $\Delta$ at resonance, the 
corresponding second set of approximately evenly spaced eigenvalues above $-1$ 
are not necessary an integer number of $\Delta$ away from the first set.

\begin{figure}[t]
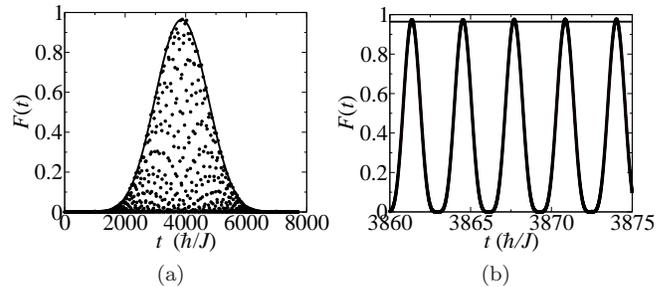

\hfill
\subfigure[]{\includegraphics[width=0.235\textwidth]{timecourse1.eps}}
\hfill
\subfigure[]{\includegraphics[width=0.235\textwidth]{timecourse2.eps}}
\hfill
\caption{The IST fidelity $F(t)$ is plotted as a function of time for $N=501$ 
at the critical impurity coupling constant $\tilde{w}\approx 8.2$. Dots 
correspond to numerical data while the solid lines correspond to the envelope 
of the output fidelity including only contributions from the critical 
eigenvectors. The slow increase in fidelity through the IST time 
$t_{\rm IST}=\pi/\tilde{\Delta}\approx 3867$ is shown in (a). A close-up of 
the behavior in the vicinity of $t_{\rm IST}$ is shown in (b). Note that the 
dots are so closely spaced here that they resemble an oscillating line, whereas
the solid line appears almost horizontal.}
\label{fig:timecourse}
\end{figure}

High-fidelity IST is in fact possible if the time is chosen to be slightly 
above or below $\pi/\Delta$. For times in the vicinity of $t\sim t_{\rm IST}$, 
Eq.~(\ref{eq:evol3}) becomes approximately
\begin{eqnarray}
\langle N|e^{-iHt/\hbar}|1\rangle&\approx&\sum_{q=1}^3\Big(
-e^{-i\lambda_2t/\hbar}\left|\langle\psi_q|1\rangle\right|^2\nonumber \\
&&\qquad + e^{i\lambda_2t/\hbar}\left|\langle\psi_q'|1\rangle\right|^2\Big).
\label{eq:ISTc}
\end{eqnarray}
Here, $|\psi_q'\rangle$ are the eigenvectors corresponding to the first 
eigenvalues above $-1$ (i.e.\ for $\lambda_q\to -\lambda_q$); note that 
$\langle 1|\psi_q'\rangle=\langle 1|\psi_q\rangle$ while
$\langle N|\psi_q'\rangle=-\langle N|\psi_q\rangle$. The right-hand side of
Eq.~(\ref{eq:ISTc}) is proportional to unity if time is chosen to be 
$t=(2n-1)\pi\hbar/2\lambda_2$, or $\tilde{t}=-\pi/2+r\pi$ where $r$ is an
arbitrary integer. For times near $t_{\rm IST}$, the probability on the output 
site varies from zero to near unity with a period $\pi\ll t_{\rm IST}$. One can 
therefore choose a time in the vicinity of $t_{\rm IST}$ at which the fidelity 
should approach unity. 

Figure~\ref{fig:timecourse} shows the IST fidelity $F(t)$, defined in 
Eq.~(\ref{eq:fidelity}), as a function of time for the particular case of
$N=501$ using the optimal impurity coupling constant 
$\tilde{w}_c\approx 8.197$. As expected, $F(t)$ reaches a maximum value at a
time near $\tilde{t}_{\rm IST}\approx 3867.44$, though the function oscillates 
rapidly throughout this slow variation. The behaviour of the fidelity including
only resonant eigenvectors in the sum~(\ref{eq:evol2}) is shown for comparison;
only the envelope of the fidelity is plotted for clarity, as this exhibits the 
same fast oscillation of the full data. Figure~\ref{fig:timecourse}(a) clearly 
shows that the time evolution of the output fidelity is governed almost 
completely by the resonant eigenvectors. A close-up of the time sequence in the 
vicinity of $t\sim t_{\rm IST}$ reveals the fast oscillation of $F(t)$; the
period is found to be $\tilde{T}_{\rm fast}\approx 3.16$ which is close to the 
predicted value of $\pi$. While $F(t_{\rm IST})\approx 0.858$, the output
fidelity actually attains a maximum $F_{\rm max}\approx 0.975$ at the slightly 
longer time $\tilde{t}=3867.70$. This maximum is attained for many 
$\tilde{T}_{\rm fast}$ periods in the vicinity of $\tilde{t}_{\rm IST}$. In 
fact, the maximum fidelity for the full dynamics slightly exceeds the value 
$F\approx 0.964$ obtained from including only the resonant eigenvectors.

\begin{figure}[t]
\centering
\includegraphics[width=0.49\textwidth, angle=0]{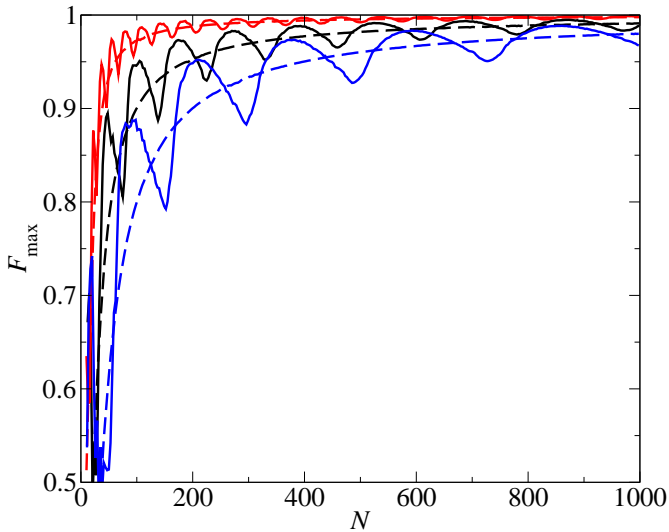}
\caption{(color online) The maximum fidelity $F_{\rm max}$ for the state 
transfer between endpoints is shown as a function of the chain length $N$. Red,
black, and blue curves correspond to $N=3m+1$, $3m$, and $3m-1$, 
$3\leq m\leq 333$, respectively. Solid lines depict the exact fidelity while
dashed lines represent the result including only the six resonant eigenvectors.}
\label{fig:fmax}
\end{figure}

The maximum fidelity $F_{\rm max}$ for quantum state transfer is shown as a 
function of $N$ in Fig.~\ref{fig:fmax} for $10\leq N\leq 1000$. The data are 
obtained by scanning the fidelity in the temporal region 
$\tilde{t}_{\rm IST}-3\tilde{T}_{\rm fast}\leq\tilde{t}\leq
\tilde{t}_{\rm IST}+3\tilde{T}_{\rm fast}$ and recording the maximum 
result for each $N$. The exact results $F_{\rm max}$, where all eigenvectors 
are included in the sum~(\ref{eq:evol2}), are shown as solid lines; the 
fidelities $F_{\rm max}'$ when the sum is restricted only to resonant 
eigenvectors is shown as dashed lines. While $N$-dependent oscillations are 
evident in the exact results, the amplitudes decrease and the wavelengths
increase with $N$; more important, though, their centers consistently follow 
the restricted fidelity curves. The restricted fidelities therefore provide an 
accurate representation of the exact fidelities in the thermodynamic limit 
$N\to\infty$. The values of restricted fidelities for large $N$ in the range 
$800\leq N\leq 1000$ are found to follow 
\begin{equation}
F_{\rm max}'\approx 
\begin{cases}
1-\frac{2.34}{N^{1.01}} & N=3m+1;\cr
1-\frac{11.4}{N^{1.04}} & N=3m;\cr
1-\frac{17.3}{N^{0.98}} & N=3m-1,\cr
\end{cases}
\end{equation}
with suitably chosen integer $m$. Thus, the IST fidelity approaches unity in
the thermodynamic limit, i.e.\ the quantum state transfer is asymptotically
perfect. According to the numerics, the state transfer error scales as 
$1-F\propto N^{-1}$ for large $N$.

\section{Discussion and Conclusions}
\label{sec:conclusions}

In this work we have shown that asymptotically perfect quantum state transfer 
is possible in uniform chains that have been modified by the addition of two
impurites, coupled to the uniform chain at the third and third-from-last sites
with strength $w$. Choosing $w\propto\sqrt{N}$, the state localized in the 
vicinity of the impurity can be tuned into resonance with chain extended
states. The associated avoided level crossing gives rise to eigenstates with 
large overlaps with the chain endpoints and with eigenvalues whose spacings
become approximately equal. The approximate linear spectrum together with 
reflection symmetry yields approximately perfect state transfer, in a time
that scales efficiently with length, as $t_{\rm IST}\propto N^{3/2}$. Indeed, 
the fidelity is found to approach unity in the thermodynamic limit 
$N\to\infty$, with error scaling as $1-F\propto N^{-1}$. To our knowledge, this 
is the only configuration with no external time-dependent or local control, 
where a uniform chain can be made to transfer quantum information perfectly in 
the limit of large system size, 

While the central insights obtained from the analytical investigations are 
validated by the explicit calculations, the numerical results reveal additional
information and display some important features. First and foremost, the 
detailed dependence of the energy splitting on $N$ at resonance (equivalently
the amplitude of the $t_{\rm IST}\propto N^{3/2}$ scaling) was only readily 
available numerically. Second, the exact time-dependence of the output 
probability was found to oscillate rapidly (with period $\pi\hbar/J$) in 
addition to the slow evolution toward maximum fidelity in the vicinity of 
$t_{\rm IST}$, independent of $N$. This means that in a practical experiment 
(with $N$ large but fixed)
the timing would have to be tested over a range of times $|t-t_{\rm IST}|\leq 
(\pi/2)(\hbar/J)$ prior to using this device to transfer unknown quantum 
information. Third, the exact value of the maximum fidelity is found to follow 
the value obtained by including only the critical eigenvectors, but for smaller
$N$ it displays pronounced oscillations. The amplitude and frequency of these
oscillations decreases steadily with $N$, so that in the thermodynamic limit
the maximum fidelity is completely dominated by the resonant eigenvectors. 

Given that the high-fidelity transfer is a direct consequence of a resonance 
between the localized and extended states, one might expect the model to be 
robust against random small errors in the chain coupling constants around $J$. 
The errors would shift the frequencies of the extended states, so that a new 
value of $w$ would need to be found to bring them back into resonance. While 
this is possible in principle, in practice finding the best value of $w$ and 
time could be difficult; if the errors are time-dependent the situation is even
worse. Unfortunately, numerical calculations suggest that the value of $F$ 
falls precipitously with noise if $w$ and $t$ are both fixed at their optimal 
noise-free value. Given $J_{i,i+1}=J\pm\delta J_{i,i+1}$ with 
random values $|\delta J_{i,i+1}|\leq x$, we find for $N=501$ that the average 
fidelity drops to $\overline{F}\sim 0.1$ for $x\approx 0.01$.

Previous studies of perfect quantum state transfer have shown that if the
evolution of a particle on a graph exhibits PST then so will its evolution on
Cartesian powers of this graph~\cite{PST3}. Because the Cartesian square and 
cube of the
uniform chain are two and three-dimensional lattices, respectively, one could 
in principle extend the current model to exhibiting PST from corner to corner 
of regular lattices in any dimension. That said, the two impurities in one
dimension translate to an unwieldy $4N$ impurities in two dimensions. Rather,
one could envisage arranging a sequence of impurities forming a half box of 
length three centered at each corner. This would ensure the presence of a 
localized state near the endpoints, which could again be tuned into resonance
through a suitable adjustment of the impurity coupling parameters. We hope to
explore this idea further in future work.

\begin{acknowledgements}
This work was supported by the Natural Sciences and Engineering Research 
Council of Canada.
\end{acknowledgements}



\bibliography{apstrefs}

\end{document}